\documentclass[%
a4paper,
twocolumn,
twoside,
superscriptaddress,
nofootinbib,
 amsmath,amssymb,
]{revtex4-1}

\usepackage{multirow}
\usepackage{graphicx}
\usepackage{dcolumn}
\usepackage{bm}
\usepackage{cases}
\usepackage[hidelinks=true]{hyperref}
\newcolumntype{P}[1]{>{\centering\arraybackslash}p{#1}}

\begin{document}


\title{Fast semidefinite programming with feedforward neural networks}

\author{Tam\'as Kriv\'achy\footnote{Corresponding author. \url{tamas.krivachy@unige.ch}}}
\author{Yu Cai}
\affiliation{%
Department of Applied Physics, University of Geneva, 
CH-1211 Geneva, Switzerland
}%
\author{Joseph Bowles}
\affiliation{ICFO - Institut de Ciencies Fotoniques, The Barcelona Institute of Science and Technology, E-08860 Castelldefels, Barcelona, Spain}
\author{Daniel Cavalcanti}
\affiliation{ICFO - Institut de Ciencies Fotoniques, The Barcelona Institute of Science and Technology, E-08860 Castelldefels, Barcelona, Spain}
\author{Nicolas Brunner}
\affiliation{%
Department of Applied Physics, University of Geneva, 
CH-1211 Geneva, Switzerland
}%

\date{\today}

\begin{abstract}
Semidefinite programming is an important optimization task, often used in time-sensitive applications. Though they are solvable in polynomial time, in practice they can be too slow to be used in online, i.e. real-time applications. Here we propose to solve feasibility semidefinite programs using artificial neural networks. 
Given the optimization constraints as an input, a neural network outputs values for the optimization parameters such that the constraints are satisfied, both for the primal and the dual formulations of the task. We train the network without having to exactly solve the semidefinite program even once, thus avoiding the possibly time-consuming task of having to generate many training samples with conventional solvers. The neural network method is only inconclusive if both the primal and dual models fail to provide feasible solutions. Otherwise we always obtain a certificate, which guarantees false positives to be excluded. We examine the performance of the method on a hierarchy of quantum information tasks, the Navascu\'es--Pironio--Ac\'in hierarchy applied to the Bell scenario. We demonstrate that the trained neural network gives decent accuracy, while showing orders of magnitude increase in speed compared to a traditional solver.
\end{abstract}

\maketitle

\section{Introduction}
Neural network-based algorithms have produced astounding results for many tasks. A notable example is that of image inpainting. Given an image with missing patches in it, properly trained algorithms can fill the patch in convincingly, matching both local and global expectations. In many fields of science and technology we find ourselves facing a similar problem: given some elements of a matrix, find a completion of it, such that the whole matrix is positive semidefinite. Such a problem is an instance of a feasibility semidefinite program (SDP), which is a widely studied class of convex optimization problems, having a broad range of applications~\cite{wolkowicz_handbook_2012,vandenberghe_applications_1999}. SDPs can be solved in polynomial time and by duality are typically able to provide a certificate of reaching the optimum. However, in various use-cases solving the problem exactly becomes impractical due to extensive runtimes, which can limit its utility in many real-time applications. This has led to several approaches to reduce the runtime, even if at the cost of losing accuracy~\cite{majumdar_recent_2020,arora_fast_2005,jain_parallel_2011,hazan_linear-time_2016,bidyarthy_toolbox_2014,mazziotti_large-scale_2011,shah_biconvex_2016}.

\begin{figure}[t]
    \centering
    \vspace*{1em}
    \includegraphics[width =0.45\textwidth]{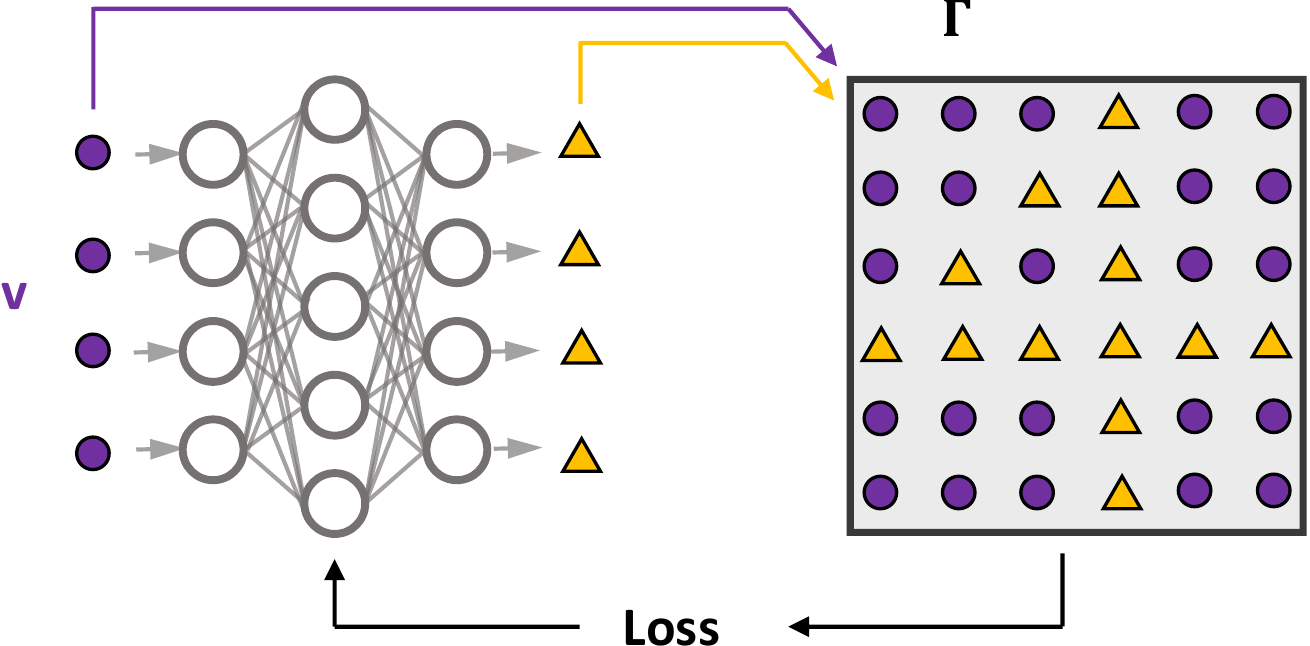}
    \caption{General scheme of our machine learning approach. Given a new instance of a feasibility SDP, described by the vector $v$ (purple dots), a neural network outputs the missing elements (yellow triangles) of the constraint matrix $\Gamma$. The neural network's parameters are updated in order to make $\Gamma$ as positive semidefinite as possible (i.e. maximize the smallest eigenvalue).}
    \label{fig:schematic}
\end{figure}

Here we explore an alternative solver for SDPs using neural networks. We replace the standard optimization procedure with a single nonlinear function, a feedforward artificial neural network. We focus on feasibility SDPs, where, given a description of the problem as an input, the neural network will guess the optimization parameter values such that they complete a matrix in a positive semidefinite way (illustrated in Fig.~\ref{fig:schematic}). If the prediction of the network results in a positive semidefinite matrix then we obtain a certificate of feasibility.
If it fails to do so, then this could be either because there does not exist a positive semidefinite completion or because the neural network just didn't find it. Luckily, the SDP has a dual formulation, in which the dual task is feasible if and only if the primal is infeasible. Thus we also check the dual task with a second neural network in a similar manner as the primal. It is only if both networks fail that we do not get a definite statement of feasibility. In other words, for the neural network-based method the duality gap is not guaranteed to be zero, while dedicated SDP solvers can in principle close this gap.

We pay this price in precision in order to gain a significantly improved runtime. Contrary to traditional solvers, for each new instance of a problem our neural network does not have to start from scratch to solve it, since it learned the structure of the problem previously, during training, which allows for a strong advantage in runtime. The flexibility of neural networks allow the user to tune this trade-off in accuracy versus runtime by changing the size and architecture of the neural network. The technique could be useful for screening data before running slower, albeit more precise solvers, when one has many instances of similarly structured optimization tasks. Additionally, it can be very useful in scenarios where quick or even real-time solving of SDPs are required, such as in calculating the safe zones of self-driving cars, collision avoidance~\cite{prajna_safety_2004,frazzoli_resolution_2012}, control systems and robotics tasks~\cite{vandenberghe_applications_1999,boyd_semidefinite_1997}, black-box quantum random number generation and quantum protocols~\cite{pironio_random_2010,brask_megahertz-rate_2017}, analyzing ground state energies in quantum chemistry and many-body physics tasks~\cite{mazziotti_first-order_2004,bravyi_approximation_2019,fukuda_large-scale_2007,nakata_variational_2001,nakata_variational_2008,navascues_paradox_2013} or bounding problems in NP~\cite{boyd_semidefinite_1997,goemans_improved_1995}.

We examine the performance of the method on a central scenario in quantum theory, Bell nonlocality~\cite{bell_einstein_1964,brunner_bell_2014}. At a glance, the basic task is to decide what kind of shared resource is necessary or sufficient for creating correlations between two parties. In particular, quantum resources, as opposed to classical ones, can lead to so-called nonlocal behavior. Previous works have used machine learning in tackling the question of whether classical resources are sufficient~\cite{krivachy_neural_2020,bharti_how_2019}, in classifying behavior by learning from many samples~\cite{canabarro_machine_2019} or by proposing new experiments~\cite{melnikov_setting_2020}. In the current work we turn towards another question: is a quantum resource sufficient to reproduce the correlations? A hierarchy of SDPs proposed by Navascu\'es, Pironio and Ac\'in (NPA) helps in answering this question~\cite{navascues_bounding_2007}. Each step of the hierarchy is a feasibility SDP which demonstrates whether the correlations under scrutiny fall within a given relaxation of the set of quantum correlations. For higher levels of the hierarchy the relaxations become tighter, and eventually converge to the true quantum set~\cite{navascues_convergent_2008, doherty_quantum_2008}. Each SDP in the hierarchy is a feasibility SDP in which some elements of the so-called moment matrix are fixed by physically observed correlations, and the other elements need to be completed such that the whole matrix is positive semidefinite.
 
We apply our technique to this problem by training two neural networks for each level of the hierarchy, one for the primal and one for the dual problem. For a fixed level of the hierarchy, the physically observed correlations define the SDP, so this is the input we use for the neural network. For training, we use only random correlations as inputs, without having to solve the SDP exactly even once. Once trained, the neural networks can predict matrix completions orders of magnitude quicker for a new correlation than conventional solvers, while keeping a decent accuracy.

In the following we introduce our approach to solving feasibility SDPs in detail. We follow by describing the Bell scenario and the hierarchy of SDPs which approximate the quantum set of correlations. Then we present the performance results and conclude with a discussion. Finally, we present a more detailed introduction to neural networks and the details of our implementation in the \emph{Methods} section.

\section{Neural networks for solving SDP}
Consider the following family of generic nonlinear optimization tasks, parametrized by a vector $v$.
\begin{align}\label{eq:generic_optimization_problem}
\nonumber \min_z \;\; &f_v(z),\\
\text{s.t.}\;\;&g_v(z) \geq 0,
\end{align}
where $f_v:\mathbb{R}^N\rightarrow \mathbb{R}$, $g_v:\mathbb{R}^N\rightarrow \mathbb{R}^M$. We focus on a differentiable objective $f_v$ and constraints $g_v$. The general spirit of our work is to encode the optimization problem in the loss function of an artificial neural network, such that it learns to satisfy the constraints and minimize the objective function, i.e. $L(v,z) = f_v(z) - \mu \min(g_v(z),0)$, where $\mu\in\mathbb{R}^+$ is introduced to balance the objective and constraints in the loss function. Then we train the neural network to minimize this loss by inputting random configuration vectors $v$, and asking the neural network to output $z$, a solution which it guesses to be optimal. After training on many random configurations from within the family the neural network will learn the structure of the problem, and when evaluated on a new instance, it will predict a close-to optimal solution. Within this family of (fixed-size) optimization problems we thus automate the solving process, since for a new instance of an optimization problem (characterized by $v$), instead of solving it from scratch, the neural network predicts the solution almost instantaneously.

In the current work we examine an important subclass of optimization problems, feasibility SDPs~\cite{wolkowicz_handbook_2012}. First, we do this because SDPs are specific instances of conic programming and thus have a dual problem, the optimal solution of which approaches the primal's from below and achieves a zero duality gap. With our technique we can train a neural network also for the dual task, and by evaluating both the dual and primal neural networks we can obtain bounds on how well the two models together manage to reduce the duality gap. Second, we only examine feasibility problems because in these the machine's task is very pronounced and its success is easy to check: predict a variable $z$ such that $g_v(z) \geq 0$. If successful, we have a certificate of feasibility. Finally, note that well-formulated SDPs with a bounded objective function can be approximated arbitrarily well with a series of feasibility SDPs using bisection techniques, with the accuracy increasing exponentially with the number of feasibility SDPs solved.

A feasibility SDP can be expressed as
\begin{align}\label{eq:primal}
\nonumber\text{Find}&\; \Gamma,\\
\text{s.t.}&\; \Gamma \geq 0,\\
\nonumber&\forall k: \langle F_k, \Gamma \rangle = v_k,
\end{align}
where $\Gamma$ is a matrix, $\langle \cdot,\cdot \rangle$ is the Hilbert-Schmidt inner product, and the matrices $F_k$ and vector $v$ encode the constraints. 
The dual task of the primal SDP can be stated as
\begin{align}\label{eq:dual}
\nonumber\text{Find}&\;\; y_k\\
\text{s.t.}&\; \begin{pmatrix}
-\sum_k y_k F_k & 0 \\
0 & \sum_k y_k v_k - \delta
\end{pmatrix} \geq 0,
\end{align}
where $\delta$ is an arbitrary positive number, which we introduced in order to be able to state the dual problem also as a feasibility task. For feasibility problems, either the primal or the dual task has a solution.

Our main tools for solving these SDPs are artificial neural networks. A feedforward neural network is a model for a generic multidimensional nonlinear function. One of its simplest realization, a multilayer perceptron, is characterized by the number of neurons per layer (width), the number of layers (depth), and the activation functions used at the neurons, which altogether model an iterative sequence of parametrized affine and fixed nonlinear transformations on the input, i.e.
\begin{align}
r_{l+1} = h(W_l r_l + b_l),
\end{align}
where the matrix $W_l$ and vector $b_l$ parametrize the linear transformation, $h$ is a fixed differentiable nonlinear function, and $r_l$ is the input of layer $l$, $r_1$ being the input of the model and $r_{\text{depth}}$ being the output. During training, the parameters of the model ($\{W_l,b_l\}_l$) are updated such that they minimize a differentiable loss function of the training set. Once trained, the neural network can be evaluated on new input instances. For a more detailed introduction to neural networks we refer the reader to the \emph{Methods} section or to Ref.~\cite{goodfellow_deep_2016}.

Our approach to solving SDPs via neural networks is illustrated in Fig.~\ref{fig:schematic}. We examine a family of feasibility SDP problems, where $\{F_k\}_k$ is fixed and only the values of $v$ are different. We input the constraints of the SDP ($v$) in a feedforward neural network. The neural network outputs the optimization variables (the free values of $\Gamma$ for the primal or $y$ for the dual) and the loss is taken to be minus one times the smallest eigenvalue of the constraint matrix. Thus the neural network tries to push the lowest eigenvalue of the constraint matrix to be as large as possible. For a test sample, if this smallest eigenvalue of the primal constraint is positive, then the machine has found a positive semidefinite completion of $\Gamma$, i.e. we obtain a certificate of feasibility. Alternatively if the dual neural network finds a positive semidefinite completion of (\ref{eq:dual}), we obtain a certificate of infeasibility of the primal task. It is only if neither the primal nor the dual neural network give positive semidefinite solutions that we are uncertain of the feasibility.

\section{Case study in quantum nonlocality}
To examine the performance of the method, we consider a hierarchy of feasibility SDPs, where the constraint matrices naturally have several elements fixed and the others left free. In the Navascu\'es--Pironio--Ac\'in (NPA) hierarchy~\cite{navascues_bounding_2007,navascues_convergent_2008}, using such SDPs we can explore the limitations of correlations coming from shared quantum resources. The hierarchy can be applied in a variety of scenarios such as estimating randomness in device independent scenarios~\cite{nieto-silleras_using_2014,bancal_more_2014}, network nonlocality~\cite{pozas-kerstjens_bounding_2019, wolfe_quantum_2020}, sequential Bell tests~\cite{bowles_bounding_2020} or bounding ground state energies~\cite{navascues_paradox_2013}, but here, for concreteness and easy benchmarking, we examine the standard bipartite Bell scenario, which is an excellent sandbox for examining the strength of correlations between two systems~\cite{bell_einstein_1964, brunner_bell_2014}.

\begin{figure}[t]
    \centering
    \vspace*{1em}
    \includegraphics[width =0.48\textwidth]{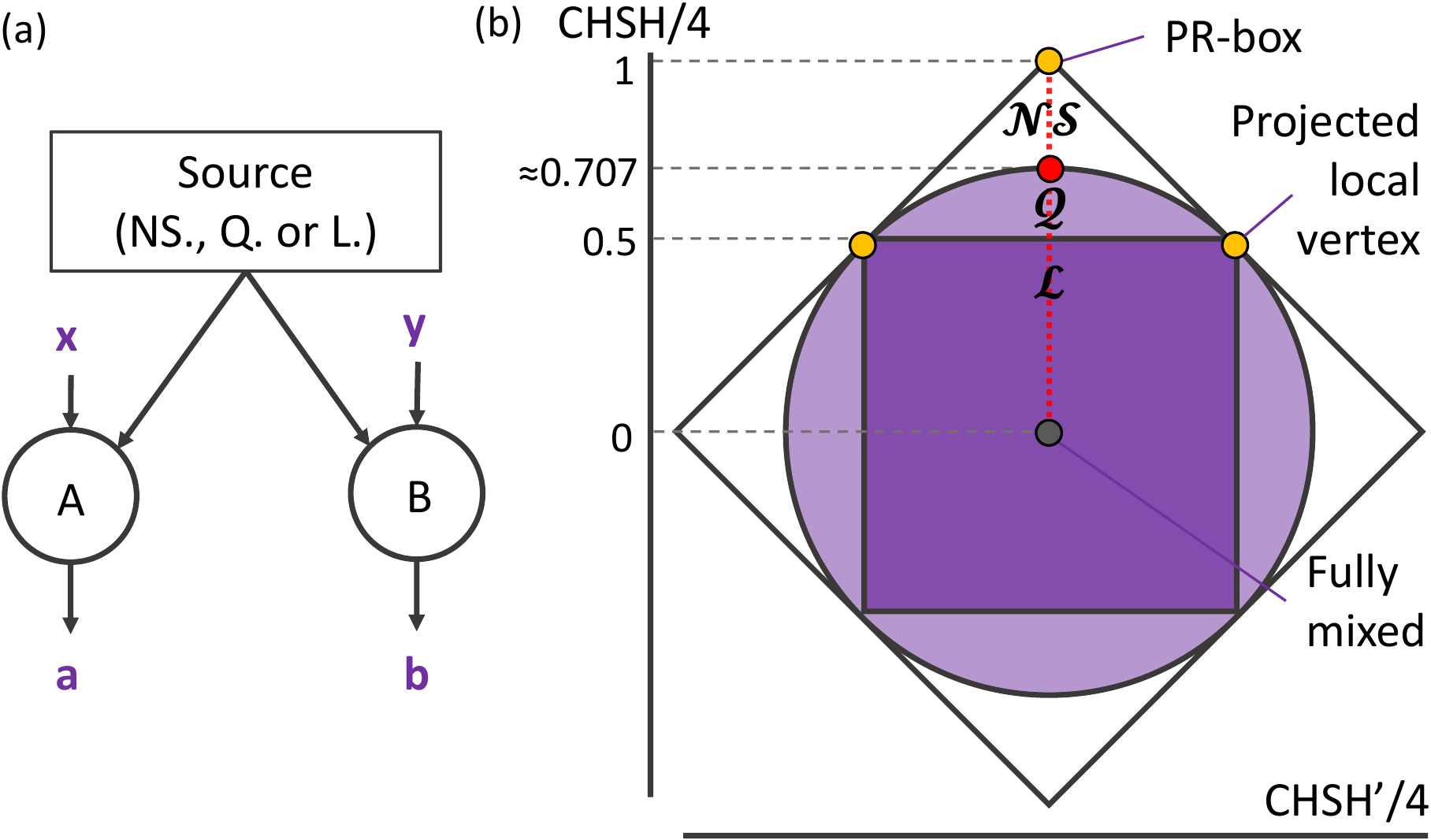}
    \caption{(a) The Bell scenario. A source (either no-signaling, quantum or local) is distributed to the two parties, Alice (A) and Bob (B). Independently from  the source they each receive and input ($x$ and $y$, resp.) and produce an output ($a$ and $b$, resp.). The correlations are characterized by the joint statistics $p(ab|xy)$. (b) A slice of the set of no-signaling correlations, where the axes are the CHSH expression divided by four and a relabeling of that (CHSH'/4). The isotropic line (dotted red line) runs between the PR-box (top yellow dot) and the fully mixed state (gray dot). The red dot represents the Tsirelson bound. The 8 local vertices achieving a CHSH value of 2 do not appear in this slice, but for illustration are projected onto 2 points (yellow dots on the border of $\mathcal{L}$).}
    \label{fig:bell-scenario}
\end{figure}

In the Bell scenario a source is distributed to two parties, Alice and Bob, and they each additionally receive an input ($x$ and $y$, respectively), and produce an output ($a$ and $b$, respectively), as illustrated in Fig.~\ref{fig:bell-scenario}(a). For the current work we will only consider discrete inputs and outputs of cardinality 2, i.e. $x,y,a,b\in\{0,1\}$. We say that there is no signaling among the parties if the statistics of the inputs and outputs obey
\begin{align}
\sum_b p(ab|xy) = \sum_b p(ab|xy') \qquad \forall a,x,y,y',\\
\sum_a p(ab|xy) = \sum_a p(ab|x'y) \qquad \forall a,x,x',y.
\end{align}
Any correlation for which these constraints hold are in the set of no-signaling correlations $\mathcal{NS}$. We enforce no-signaling and that the inputs are independent from the source. Then, just by observing the statistics of the inputs and the outputs of the two parties, $p(ab|xy)$, one can in certain cases deduce that stronger than classical sources were used, e.g. quantum or post-quantum sources. With any classical source, the obtained statistics can be described by a shared random variable $\Lambda$, constraining the correlations to be of the form
\begin{align}
\label{eq:lhv}
p(ab|xy) = \int d\lambda p(\lambda) p(a|x, \lambda) p(b|y, \lambda),
\end{align}
for some $p(a|x,\lambda), p(b|y,\lambda)$ conditional probability distributions, otherwise known as response functions, and some probability distribution $p(\lambda)$. Any correlation which can be described by such a model (otherwise known as a local hidden variable model) is within the local set $\mathcal{L}$.

The Born rule of quantum theory postulates that for any quantum source, $p(ab|xy)$ must have the form
\begin{align}\label{eq:quantumBorn}
p(ab|xy) = \text{Tr}(\rho A_x^a \otimes B_y^b),
\end{align}
where $\rho$ is a Hermitian positive semidefinite matrix (or operator, more generally) with unit trace, and can be thought of as the quantum analogue of a probability distribution, $\{A_x^a\}_{a,x}$ are projective matrices such that $\forall x:\,\sum_a A_x^a = \mathbb{I}$, and similarly for $B_y^b$. The set $\{A_x^a\}_a$ represents Alice's measurement when her input is $x$ and output is $a$, and analogously for Bob. Any correlation having a quantum explanation according to the above equation is in the set $\mathcal{Q}$.

The no-signaling set and the local set are polytopes~\cite{brunner_bell_2014}, bounded by convex combinations of a finite set of vertices, or equivalently by linear inequalities. For the local set there are 8 non-trivial inequalities, all of them equivalent up to relabeling to the same Clauser--Horne--Shimony--Holt (CHSH) inequality, which is a linear expression bounded by the value $2$ for local behaviors~\cite{clauser_proposed_1969}. This inequality defines a facet, spanned by 8 local vertices. The no-signaling set is bounded by 24 vertices of which 16 are also local vertices, and 8 are genuinely no-signaling vertices, all equivalent under relabeling to the Popescu--Rohrlich box (PR-box)~\cite{popescu_quantum_1994, rastall_locality_1985, khalfin_quantum_1985,barrett_nonlocal_2005}, defined as
\begin{equation}\label{eq:PRbox}
p^{\text{PR}}(ab|xy) := 
\begin{cases}
    1 & \text{if } a\oplus b = xy,\\
    0              & \text{otherwise},
\end{cases}
\end{equation}
which gives a CHSH value of $4$, far above the local bound. The relation between $\mathcal{L},\mathcal{Q}$ and $\mathcal{NS}$ is visualized on a two-dimensional slice of the space in Fig.~\ref{fig:bell-scenario}(b).

Quantum behaviors can at most achieve a CHSH value of $2\sqrt{2}$, the Tsirelson bound, thus achieving nonlocality~\cite{cirelson_quantum_1980}. However, the quantum set is not a polytope and is thus more difficult to characterize than the local or no-signaling sets. An additional difficulty is that one can not in general bound the dimension of the density matrix required. A general technique for approximating the quantum set is an infinite hierarchy of relaxations, known as the NPA hierarchy~\cite{navascues_bounding_2007}. Each relaxation is defined by a finite set of strings of moments of the quantum measurement operators of Alice and Bob. In this paper we will examine the relaxations $\mathcal{Q}1$, $\mathcal{Q}1$AB, $\mathcal{Q}2$, $\mathcal{Q}3$, where the number denotes the maximal degree of moments used in the relaxation, and for $\mathcal{Q}1$AB only includes single moments and cross-moments of Alice's and Bob's measurement, a commonly used intermediate level between $\mathcal{Q}1$ and $\mathcal{Q}2$. The set $\mathcal Qn+1$ is a subset of $\mathcal Qn$, and as $n$ approaches infinity we are guaranteed to recover the actual quantum set $\mathcal{Q}$~\cite{navascues_convergent_2008,doherty_quantum_2008}.

The basic idea of the relaxations is that if a distribution has a quantum description according to Eq.~(\ref{eq:quantumBorn}), then the so-called moment matrix should be positive semidefinite. A moment matrix, $\Gamma$, is constructed by taking a set of strings of moments of measurement operators, e.g. for $\mathcal{Q}1$ one takes $S:=\{\mathbb{I}, A_0^0, A_1^0, B_0^0, B_1^0\}$. Other single moments are not needed due to normalization, so for clarity we leave away the output superscript from now on. One constructs the moment matrix by taking the expectation value of the dyadic product of $S$ and its element-wise adjoint set $S^\dag$, which in the case of single moments is the same as $S$ since the projectors are self-adjoint. The moment matrix is the following, where the $\langle \cdot \rangle$ is the expectation value with respect to $\rho$, i.e. $\text{Tr}(\rho \,\cdot)$. Only the upper triangle is shown as $\Gamma$ is symmetric.
\begin{align}\label{eq:gamma}
\begin{tabular}{c|ccccc}
 & $\mathbb{I}$ & $A_0$ & $A_1$   & $B_0$ & $B_1$   \\
\hline \textbf{$\mathbb{I}$} & $\langle \mathbb{I}\rangle$  & $\langle A_0\rangle$  & $\langle A_1\rangle$  & $\langle B_0\rangle$    & $\langle B_1\rangle$            \\
$A_0$ &             & $\langle A_0\rangle$          & $\langle \mathbf{A_0A_1}\rangle$ & $\langle A_0B_0\rangle$        & $\langle A_0B_1\rangle$          \\
$A_1$ &             &             & $\langle A_1\rangle$            & $\langle A_1B_0\rangle$        & $\langle A_1B_1\rangle$          \\
$B_0$ &             &             &               & $\langle B_0\rangle$          & $\langle \mathbf{B_0B_1}\rangle$ \\
$B_1$ &             &             &               &             & $\langle B_1\rangle$           
\end{tabular}
\end{align}

For any behavior derived from some quantum source $\rho$, such a moment matrix must be positive semidefinite. Notice that even without knowing $\rho$, almost all elements can be identified with the observed statistics $p(ab|xy)$ via the Born rule in Eq.~(\ref{eq:quantumBorn}). However, a few elements are unphysical and can not be observed, since they would require the joint measurement of two of Alice's or Bob's operators, which is prohibited for noncommuting measurements. For $\mathcal{Q}1$ there are only two such unknown elements, denoted by bold text in the matrix (\ref{eq:gamma}), however for higher levels their number grows quickly, namely there are 8, 22, 52, 92 and 142 unknown elements in the moment matrices of levels $\mathcal{Q}1$AB$, \mathcal{Q}2,\dots \mathcal{Q}5$, while each of them have the same number of elements fixed by the statistics $p(ab|xy)$.

The task is now the following: given some statistics from a Bell scenario, $p(ab|xy)$, and a level of the hierarchy (say $\mathcal{Q}1$), find real numbers for the unknown elements of the moment matrix such that it is positive semidefinite. If such a completion is possible then $p(ab|xy)\in\mathcal{Q}1$. We can see that this problem is of the form (\ref{eq:primal}) by defining the matrices $\{F_k\}_k$ to have ones only in the matrix indices where physical elements appear and zero otherwise, and by identifying the entries of $v$ with the appropriate elements of $p(ab|xy)$. As such, a dual formulation also exists, such that is if there is a positive semidefinite completion of the dual constraint matrix (see Eq.~(\ref{eq:dual})), then we certify that $p(ab|xy)\notin\mathcal{Q}1$. Otherwise, if we do not find a positive semidefinite for either the primal or dual, then we are left in uncertainty. In the following we train neural networks for the primal and dual of this problem for different levels of the NPA hierarchy and examine their performance. 

We examine two ways to generate input behaviors $p(ab|xy)$. Since local behaviors can easily be screened out by checking their CHSH values, we do not need to train the machine to perform well on those. Thus we take samples from the nonlocal part of the no-signaling set, $\mathcal{NS}\backslash \mathcal{L}$. We examine only a non-redundant part of this set by not considering the 8-fold symmetry under relabeling. As a result the set we consider is spanned by the 8 extremal local points and the PR-box which maximize the canonical CHSH inequality (see the yellow dots in Fig.~\ref{fig:bell-scenario}(b) for an illustration). With a slight abuse of notation we continue to label this set as $\mathcal{NS}\backslash \mathcal{L}$.

In the first sampling technique, we generate random samples from $\mathcal{NS}\backslash \mathcal{L}$ using \emph{hit and run sampling}~\cite{smith_efficient_1984, belisle_convergence_1998}. In the second, which we will refer to as \emph{weighted vertex sampling} we take uniformly randomly weighted mixtures of the 9 vertices of $\mathcal{NS}\backslash \mathcal{L}$, with 8-fold weight on the PR-box. Formally, if the local vertices are $\{p^\mathcal{L}_i(ab|xy)\}_{i=1}^8$ and the PR-box is $p^{\text{PR}}(ab|xy)$, then a training sample is generated as
\begin{equation}\label{eq:weighted_vertex}
p(ab|xy) = \frac{8 w_0\, p^{\text{PR}}(ab|xy) + \sum_{i=1}^8 w_i\, p^\mathcal{L}_i(ab|xy)}{8w_0 + \sum_{i=1}^8 w_i},
\end{equation}
where the $w_i\in[0,1]$ are uniformly drawn random numbers. This sampling technique resulted in a more reliable training for the dual neural network.

Among other factors, the accuracy of the trained network depends on the architecture of the neural network, in particular on its size. In the following we will examine a family of neural networks, one for each level of the hierarchy. We keep the depth at a constant of 8 layers, while we vary the width in order to accompany the growing complexity of higher levels. In particular we take the width to be three times the number of elements which need to be predicted. The last layer is different from the preceding ones, first off, because its width is exactly the size of the number of elements which need to be predicted. Second, whereas in other layers we use exponential linear activations, we use a linear activation for the final layer of the primal network. For the dual network using linear activation can lead to instability, since the optimal solution to $y_k$ for maximizing the eigenvalue of the dual matrix in Eq.~(\ref{eq:dual}) can tend to infinity. Thus we limit the outputs such that $-1 < y_k < 1$ by using a tangent hyperbolic activation. For a small enough $\delta$ (e.g. we used $3e-7$) this only minimally restricts the generality of the solver, giving errors well below the standard inaccuracy of the machine learning predictions. For additional details of the training procedure see the \emph{Methods} section.

\section{Results}
\begin{figure}[t]
    \centering
    \vspace*{1em}
    \includegraphics[width =0.5\textwidth]{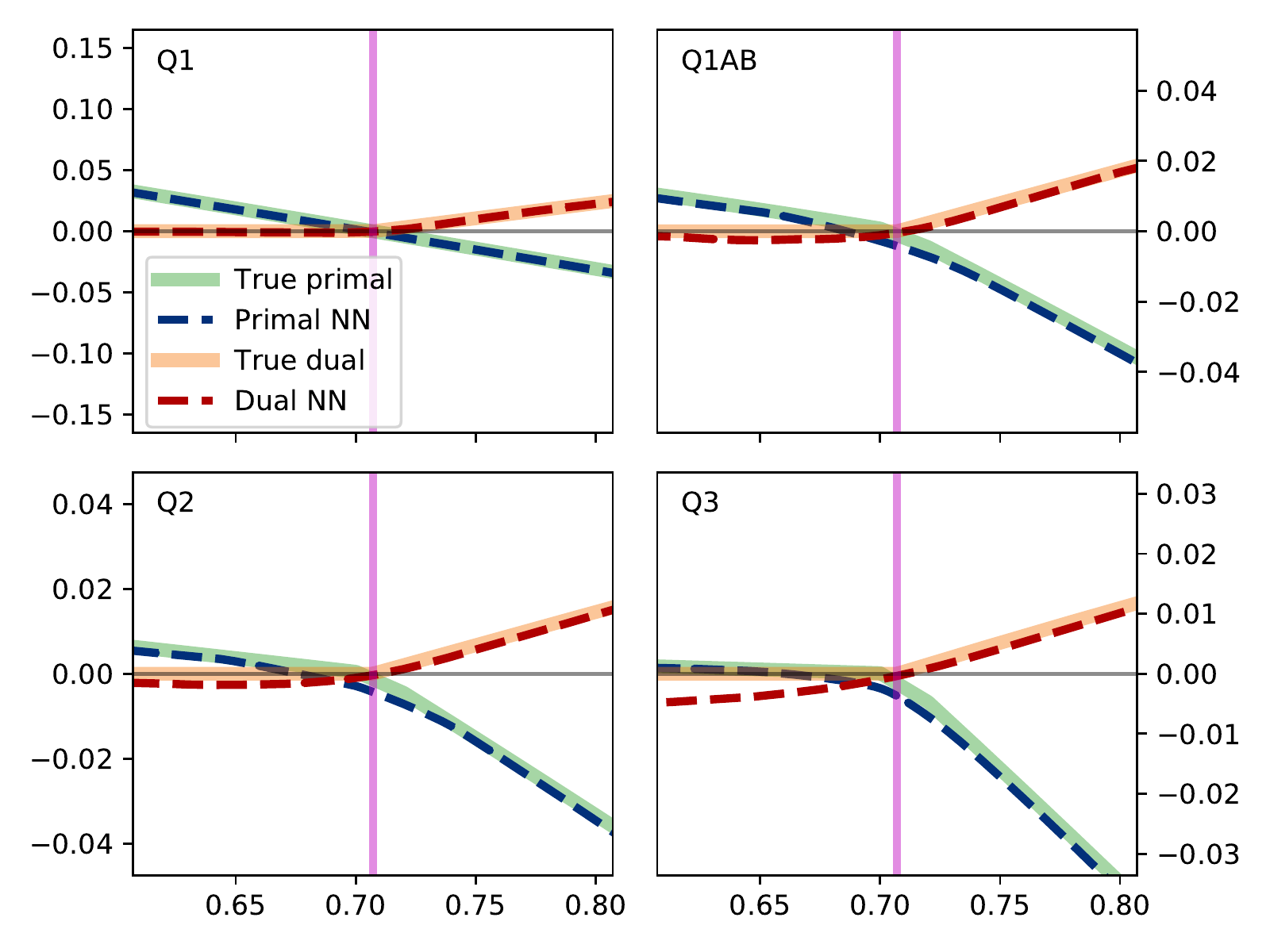}
    \caption{The maximum of the smallest eigenvalue of the primal and dual SDP matrix as predicted by the machine (blue and red dashed lines, resp.), compared to the true value calculated by an SDP solver (green and orange solid lines, resp.), as a function of the mixing parameter $q$ of the isotropic line, for several different levels of the hierarchy. For the dual methods, in order to have numerical stability the outputs are restricted to be between $-1$ and $1$. The Tsirelson bound at $q=1/\sqrt{2}$ is denoted by the solid magenta line.}
    \label{fig:tsirelson}
\end{figure}

The memory usage of the neural network method is approximately the same as for conventional SDP solvers, with the construction of the moment matrix being the primary bottleneck. Hence, in the following we compare the two approaches from two other aspects, their accuracy and runtime.

First let us examine Tsirelson's bound in detail by using both the primal and the dual neural networks for relaxations $\mathcal{Q}$1, $\mathcal{Q}$1AB, $\mathcal{Q}$2 and $\mathcal{Q}$3 on the isotropic line, i.e. on distributions
\begin{equation}
p_q(ab|xy) = q \,p^{\text{PR}}(ab|xy) + (1-q)\,p^{\text{id}}(ab|xy),
\end{equation}
where $p^{\text{id}}(ab|xy)$ is the completely flat distribution and $q\in[0,1]$ is the mixing parameter. The edge of the quantum set, Tsirelson's bound, is at precisely $q^*=\frac{1}{\sqrt{2}}$. The isotropic line is portrayed in Fig.~\ref{fig:bell-scenario}(b). In Fig.~\ref{fig:tsirelson} we depict the minimum eigenvalue of the predicted moment matrix as a function $q$, and contrast it to the maximum possible minimum eigenvalue, calculated by an SDP solver. For the results in Fig.~\ref{fig:tsirelson} we use hit and run training for the primal networks and weighted vertex for the duals, since the dual neural network showed poor convergence and performance when trained on hit and run sampling.

We further explore the effect of choosing one training sampling technique versus the other by evaluating the accuracy of the $\mathcal{Q}$2 primal neural network solver for both hit and run and weighted vertex sampling. When testing the performance we additionally included a set of correlations coming from random two-qubit pure states measured in random projective bases (all sampled uniformly in the Haar measure) as well as results from the isotropic line. Results are portrayed in Table~\ref{table:Q2}.

\begin{table}
  \centering
  \renewcommand{\arraystretch}{1.2}
  \begin{tabular}{|P{2.5cm}|P{1.7cm}|P{1.7cm}|}

    \hline
    \textbf{Tested on} & \multicolumn{2}{c|}{\textbf{Trained on}} \\
     & Hit and run & Weighted vertices\\
    \hline
    Hit and run & \textbf{81\%} & 60\% \\ \hline
    Weighted vertices & \textbf{77\%} & \textbf{77\%} \\ \hline
    Quantum pure 	& \textbf{80\%} & 50\% \\ \hline
    Tsirelson bound (nonlocal part) & 0.176/0.207 = 85\% & 0.193/0.207 = \textbf{93\%}  \\ \hline
  \end{tabular}
  \caption{Percent of test samples for which the $\mathcal{Q}$2 primal neural network found a positive semidefinite completion, for different training and test sampling techniques. For the test sets, consisting of 10000 samples each, we used only correlations which have positive semidefinite completions. For the Tsirelson bound the percentage is calculated only on the nonlocal part of the isotropic line, i.e. $q=0.5$ is $0\%$ and $q=\frac{1}{\sqrt{2}}$ is $100\%$.}
  \label{table:Q2}
\end{table}

Finally, in Table~\ref{table:timings} we compare the time performance of the primal neural network to standard solving software (MOSEK), solved on the same personal computer at 100\% CPU usage. We note that the timing results depend on the choice of neural network architecture. Recall that for different levels of the hierarchy we kept the depth fixed and increased the width linearly. This was an arbitrary choice which leads to the accuracy results, also visible in Table~\ref{table:timings}. If one would like to have higher accuracy on larger problem sizes, one could use larger neural networks, at the cost of having somewhat slower evaluation times. We can see that for the choice we made the speed of solving is orders of magnitude quicker than solving with MOSEK, hence there is still much space left for improving the accuracy if required.

\begin{table}
  \centering
  \renewcommand{\arraystretch}{1.2}
  \begin{tabular}{|P{1.7cm}|P{1.3cm}|P{1.3cm}|P{1.3cm}|P{1.3cm}|}
    \hline
    & Q1 & Q1AB & Q2 & Q3 \\
    \hline
    SDP variables & 2  & 8 & 22   & 52\\ \hline
    NN accuracy & 98\%  &  79\% & 81\% & 60\%\\ \hline
    NN Tsirelson & 99\% & 97\%  & 85\% & 77\%\\ \hline
    NN time per sample & 17 $\mu$s  &  20 $\mu$s & 36 $\mu$s & 210 $\mu$s\\ \hline
    MOSEK time per sample & 9 ms & 10 ms & 13 ms & 17ms\\ \hline
  \end{tabular}
  \caption{Time per sample and accuracy for different levels of the hierarchy. The neural networks (NN) used all have a depth of 8 and a width which is 3 times the number of SDP variables which need to be predicted.}
  \label{table:timings}
\end{table}

\section{Discussion}
We have developed a solver for feasibility SDPs of a fixed structure using feed-forward neural networks. Our approach is an unsupervised learning one in the sense that training is done only with random input samples, and the neural network must learn itself the best output, which gives ``the most positive-semidefinite'' constraint matrix by constructing it explicitly. This is facilitated through a loss function which motivates the neural network to maximize the smallest eigenvalue of the generated constraint matrix. In future work it could be interesting to explore alternative approaches more in line with supervised learning, where one generates a training set by taking many random inputs and calculating the solution using conventional solvers, which input-output pairs would be training data for a neural network. Though it could be more precise than the approach explored in this work, it comes with the disadvantage that generating training samples is slow. This is a similar bottleneck as for the training of the method examined in this work, since we must calculate the eigenvalues of the constraint matrix for each training sample, which calculation for an $m \times m$ matrix typically scales as $m^3$ in practice. We note that we have tried briefly to do a supervised learning approach in which we circumvent these timely training steps by generating random positive semidefinite matrices and asking a feedforward neural network to predict the missing element positions given the known element positions. This gives a quick training procedure, however it is not as accurate as the technique demonstrated here, since the learned completions are not trying to push the matrix to be ``as positive semidefinite as possible''.

A direction worth exploring in future work would be the use of generative methods. Instead of learning the distribution over outputs given an input (as it is done in discriminative techniques), generative neural networks learn the distribution over the inputs, and can generate new instances. Though we were inspired by such generative machine learning-based image inpainting, the task here is to always predict \emph{the same} unknown elements of the matrix given the constraints $v$. Thus it is unnecessarily expensive to learn a generative distribution over all positive semidefinite matrices, when one only needs to learn the conditional probabilities of some elements given others. Nonetheless, if sufficiently strong generative techniques were used their performance could be comparable to ours, though the same trade-off between training speed and accuracy is expected to also be present for generative models.

Examining generative models lead us towards more general scenarios than the one considered here. In the current work we discussed families of feasibility SDP problems where the structure of the problem is fixed. This is a common scenario in applications, where $\{F_k\}_k$ (the structure of the problem) is essentially fixed, and the task must be solved for different specific instances given by $v$. If this is not the case then in principle $F_k$ can be provided next to $v$ to the neural network. For the example of matrix completion, a different ${F_k}_k$ would mean that different elements of the matrix need to be filled in, which for example a generative model could tackle, as previously discussed. In the current work we evaluated our approach on such tasks, i.e. tasks where the structure of the problem requires us to fill in a matrix, as shown in Fig.~\ref{fig:schematic}, where the entries of $\Gamma$ are either completely fixed by $v$ or are free optimization variables. For more general settings variable elimination on the constraints can help reduce the problem to this form, then allowing for our method to be used with slight modifications.

Furthermore, one could examine even more generic tasks by moving outside the realm of semidefinite programs to generic nonlinear objective functions and constraints, as in Eq.~(\ref{eq:generic_optimization_problem}), where neural networks could be trained to minimize a Lagrangian function, incorporating all constraints and objectives with penalty terms. There have been many works addressing optimization problems using machine learning~\cite{bengio_machine_2020}, however previous works addressing SDP specifically have been mostly focused on developing neural networks for specialized hardware. In principle these are quick, and for specific problems, such as for semidefinite programming, they can guarantee convergence~\cite{danchi_jiang_recurrent_1999, hopfield_neural_1985}. However, if simulated on a general-purpose digital computer, these neural network optimizations are also slow, since they must evaluate the constraints many times before converging.

The choice of the training set is important, especially since we are working in high-dimensional spaces. For example training on random quantum points for the primal is very inefficient, as most of them are local points. Another example is our preference of using the weighted vertex sampling for the dual neural network. Most samples generated with hit and run sampling are within $\mathcal{Q}$, so the dual network focuses on performing well on that region. However, in order to function well on the isotropic line, for example, it must also perform well on the region outside $\mathcal{Q}$, so that it can detect the boundary. We overcame this difficulty by introducing the weighted vertex sampling. In general some domain knowledge is useful in choosing the training set for such neural network-based SDP solvers.

Finally, an interesting question worth exploring more in depth is the advantage of neural networks versus conventional solvers. In which scenario could such a neural network approach have an advantage compared to standard solvers? Clearly, in application where many instances of semidefinite programs must be solved quickly, a neural network-based approach could be useful, for example in calculating the safe zones of self-driving cars, collision avoidance~\cite{prajna_safety_2004,frazzoli_resolution_2012}, control systems and robotics tasks~\cite{vandenberghe_applications_1999,boyd_semidefinite_1997}, black-box quantum random number generation and quantum protocols~\cite{pironio_random_2010,brask_megahertz-rate_2017}, testing many different candidate materials in quantum chemistry and many-body physics tasks~\cite{mazziotti_first-order_2004,bravyi_approximation_2019,fukuda_large-scale_2007,nakata_variational_2001,nakata_variational_2008,navascues_paradox_2013} or bounding problems in NP~\cite{boyd_semidefinite_1997,goemans_improved_1995}. Furthermore, we note that there are additional potential advantages of using neural network solvers for SDPs. One can use standard neural network analysis tools, such as optimizing over the input to see which inputs trigger the strongest response of the network. For example in the NPA hierarchy if the distribution $p_\varphi(ab|xy)$ depends on parameters $\varphi$ in some nonlinear manner, then SDP solvers could not help us. However, with a trained neural network we could ask it to find the a $\varphi^*$ for which the maximal smallest eigenvalue is \emph{as negative} as possible, thus to find ``very nonlocal'' behaviors in a sense.

In conclusion, due to the rich variety of machine learning architectures, there could be a myriad of ways to tackle semidefinite programming. Here we demonstrated a simple feedforward neural network approach to optimization with semidefinite constraints which, by construction, is unable to give false positives. The approach is also applicable to the dual formulation of the SDP. Our neural network, once trained, is orders of magnitudes quicker than standard SDP solvers when evaluating multiple instances of the same type of problem. Though it is less precise, the performance is acceptable for the archetypal Bell scenario, and by changing network architectures, one can tune the trade-off of precision and speed.

\section{Methods}\label{sec:methods}
Here we introduce a simple artificial neural network architecture, the multilayer perceptron, and the details of our implementation for feasibility SDP solving. For a more in-depth, pedagogical introduction to modern neural networks we refer the interested reader to Ref.~\cite{goodfellow_deep_2016}.

A multilayer perceptron is a model for a generic nonlinear multivalued, multivariate function, i.e. $H_{\text{NN}} : \mathbb{R}^s \rightarrow \mathbb{R}^t$. It is constructed of a series of layer, each layer consisting of an affine and a subsequent nonlinear transformation. The depth of a neural network is the number of layers used, $d$. A layer $l$ is characterized by its width $w_l$ and the activation function used in it, $h_l$, though typically all layers (except the first and the last) have the same width $w$ and activation $h$. The output of layer $l$ is 
\begin{align}
r_{l+1} = h_l(W_l r_l + b_l),
\end{align}
where the weight matrix $W_l\in \mathbb{R}^{w_{l+1}} \times \mathbb{R}^{w_l}$ and bias vector $b_l \in \mathbb{R}^{w_{l+1}}$ parametrize the affine transformation, $h_l$ is the fixed differentiable nonlinear ``activation'' function, and is typically an element-wise operation, and $r_l\in \mathbb{R}^w$ is the input of layer $l$, $r_1 \in \mathbb{R}^{s}$ being the input of the model and $r_{d}\in\mathbb{R}^t$ being the output. The free, trainable parameters of the model $\{W_l, b_l\}_l$ are often collectively referred to as the weights and are denoted by $\theta$.

A neural network is trained on the so-called training data $R=\{r_1^i\}_i$, with the objective of minimizing a loss function $\mathcal{L}(R|\theta) = \sum_i L(r_1^i,r_d^i | \theta)$. The most common algorithm to train a neural network is stochastic gradient descent, in which a random subset of the training data, a batch $R_j$, is taken and evaluated by the neural network. For each batch the loss is evaluated and the weights are updated according to the gradient of the loss function,
\begin{align}
\theta' = \theta - \eta \nabla_\theta \mathcal{L}(R_j|\theta),
\end{align}
where $\eta\in\mathbb{R}^+$ is the learning rate. The process is repeated with all batches of the data set. If the data is finite, then the algorithm reiterates over the data for many epochs. However, in the application analyzed in this work new training data can be artificially generated on the go, hence we do not use epochs, but use fresh data for each round. There are many variants of stochastic gradient descent. In the current work we additionally applied momentum to the updates, i.e. the updater remembers the update of the previous step, $\Delta \theta$, and adds this with a weight $\alpha\in\mathbb{R}^+$, such that
\begin{align}
\theta' = \theta - \eta \nabla_\theta \mathcal{L}(R_j|\theta) + \alpha \Delta \theta.
\end{align}
This is an example of a procedure which helps to avoid getting stuck in local minima during training. Typically the learning rate is decayed during training so that we can fine-tune the model towards the end.

Training is typically done until the loss appears to converge or until a fixed number of steps. Once trained, the neural network can be evaluated on new input instances, and is often benchmarked on a test set, which is composed of data that it hasn't seen during training.

In the current work we used neural networks of depth 8 and a width is 3 times the number of SDP optimization parameters (see Table~\ref{table:timings} for primal task sizes). The final layer width is precisely the number of SDP optimization parameters. The input size is always 8, as this is the number of parameters needed to characterize the conditional probability vector $p(ab|xy)$ for binary variables (i.e. the number of unique, physically observable elements in (\ref{eq:gamma})). The loss used is
\begin{align}
L(r_1,r_d | \theta) = - \min\,\text{e.v.}\left(\Gamma(r_1, r_d|\theta)\right),
\end{align}
where we introduced the function $\Gamma(\cdot)$, which creates the constraint matrix from the inputs and outputs (see Eqs.~(\ref{eq:primal}), (\ref{eq:gamma}) for the primal and Eq.~(\ref{eq:dual}) for the dual), and the $\text{e.v.}(\cdot)$ function which returns the set of eigenvalues of a matrix. Such an eigenvalue function is implemented in the TensorFlow machine learning library~\cite{martin_abadi_tensorflow_2015}. Note that the dependence on the weights $\theta$ is implicitly in the output $r_d$. Also, recall that the input $r_1$ corresponds to $v$ in the generic introduction in the maintext, or $p(ab|xy)$ in the examined quantum case study.

Before feeding any input correlation into the network (be that training or testing) we permute the labels in $p(ab|xy)$ such that it maximally violates the canonical CHSH inequality (see e.g. Ref.~\cite{brunner_bell_2014}), then we apply the standard preprocessing used with neural networks, that for each feature the mean is 0 and variance is 1. In each round of training we generate 10000 random $p(ab|xy)$ vectors either through hit and run sampling of the no signaling polytope or with the weighted vertex sampling (see Eq.~(\ref{eq:weighted_vertex})). We perform 800 rounds of training in total with stochastic gradient descent, of which the first 200 rounds have a fixed learning rate of 0.005 (0.0001 for the dual network), while the rest have a decay rate of 0.99 per round, and the momentum is kept at a constant rate of 0.8.

The neural network used for the dual task (Eq.~\ref{eq:dual})) is slightly different from the one used for the primal. We have to restrict the outputs to be finite, since the optimal solution tends to infinity for points where the dual is feasible; we do this by replacing the linear activation used in the final layer of the primal neural network with a tangent hyperbolic activation function. All other activations in the networks are exponential linear functions. Even with this restriction the dual neural network tries to increase its weights as much as possible in order to get marginal improvements in the tangent hyperbolic functions (pushing outputs of neurons as close to 1 as possible). This can lead to exploding gradients, which we counteract by using the default Keras L2 activity regularizer on each layer with weight $10^{-6}$~\cite{chollet_keras_2015}.

\section{Acknowledgments}
Most of the computations were performed at University of Geneva on the Baobab cluster. TK, YC and NB acknowledge financial support from the Swiss National Science Foundation (Starting grant DIAQ, and QSIT), and the European Research Council (ERC MEC). 
JB and DC acknowledge support from Fundaci\'o Cellex, Fundació Mir-Puig, Generalitat de Catalunya (CERCA, AGAUR SGR 1381). JB additionally acknowledges support from the Spanish MINECO (Severo Ochoa SEV-2015-0522) and the AXA Chair in Quantum Information Science. DC additionally acknowledges support from the Government of Spain (Ramon y Cajal fellowship, FIS2020-TRANQI and Severo Ochoa CEX2019-000910-S) and ERC AdG CERQUTE.
\bibliography{manuscript}
\end{document}